\documentstyle[abe,NP,12pt]{article}
\begin{document}
\input mssymb.tex
\pagestyle{empty}
\baselineskip20pt
\def\today{\ifcase\month\or
  January\or February\or March\or April\or May\or June\or
  July\or August\or September\or October\or November\or December\fi
  \ \ \number\year}
\vspace*{-40pt}
\rightline{\bf RIMS-1161}
\vspace*{30pt}
\centerline{\cmssB Proof of the Gauge Independence}
\vskip10pt
\centerline{\cmssB of the Conformal Anomaly of Bosonic String}
\vskip10pt
\centerline{\cmssB in the Sense of Kraemmer and Rebhan}
\vskip50pt
\centerline{
 Mitsuo Abe$^{\hbox{\sc a},\,}$\foot(1,{
  E-mail: abe@kurims.kyoto-u.ac.jp}) 
 and 
 Noboru Nakanishi$^{\hbox{\sc b},\,}$\foot(2,{
  E-mail: nbr-nakanishi@msn.com})}
\vskip15pt
\centerline{\it $^{\hbox{\sc a}}$Research Institute for Mathematical Sciences,
Kyoto University, Kyoto 606-01, Japan}
\vskip5pt
\centerline{\it $^{\hbox{\sc b}}$12-20 Asahigaoka-cho, Hirakata 573, Japan}
\vskip20pt
\vskip100pt
\centerline{\bf Abstract}
Kraemmer and Rebhan claimed the gauge independence of the conformal 
anomaly of bosonic string for various gauge fixings in the framework 
of the perturbation theory of two-dimensional quantum gravity.
It is pointed out that their proof is wrong.  The gauge independence 
is proved for the gauge-fixings which reduce to the linearized de Donder 
gauge in the flat limit of the background metric.
Similar remarks are made also for the Rebhan-Kraemmer current anomaly.
\vskip20pt
\noindent{{\it PACS:\/} 11.25.-w; 04.60.Kz}\hfill\break
\vfill\eject
\pagestyle{plain}
\setlength{\oddsidemargin}{.5truecm}
\setlength{\textheight}{23.cm}  
\setlength{\textwidth}{16.cm}
\setlength{\topmargin}{-.5cm}
\setlength{\baselineskip}{19.8pt}
\setlength{\parindent}{25pt}
\textfont0=\tenrm  \textfont1=\teni \textfont2=\tensy \textfont3=\tenex
\def\rm{\fam0 \tenrm} \def\mit{\fam1 } \def\cal{\fam2 }
\def\bf{\tenbf}  \def\it{\tenit} \def\sl{\tensl}
\scriptfont0=\sixrm  \scriptfont1=\sixi  \scriptfont2=\sixsy
\scriptscriptfont0=\smallr \scriptscriptfont1=\smalli 
                           \scriptscriptfont2=\smallsy
\rm
\Sec{Introduction}
As is well known, the bosonic string can be described by the two-dimensional
quantum gravity coupled with $D$ scalar fields, where $D$ denotes 
the dimension of the world where a string lives.
In noncovariant gauges such as the conformal gauge, the conformal anomaly
disappears if and only if $D=26$.
\par
About one decade ago, several authors [1--6]
extended this result to the case of covariant gauges.  
Their method was to calculate the conformal-anomaly term in the two-point 
function of ``energy-momentum tensor'' $T_{\mu\nu}$ in the framework of 
perturbation theory.  All of them claimed that the critical dimension 
$D=26$ was obtained.
\par
A remarkable proposition, which we call KR proposition, was put forward by
Kraemmer and Rebhan \cite{Kraemmer-Rebhan}: They claimed the {\it gauge 
independence of the conformal anomaly\/}.  They gave a very simple proof of 
it based on the BRS invariance only.
\par
On the other hand, in 1992, the present authors \cite{AN1} explicitly 
demonstrated the indefiniteness of the definition of $T_{\mu\nu}$ in
the de Donder gauge and pointed out that the conformal-anomaly term depends
on the choice of $T_{\mu\nu}$ so that $D=26$ is not necessarily obtained.
\par
Since the above two claims look apparently contradictory, we have examined
the KR proposition closely.  Although many explicit examples support the KR
proposition, we have found that the proof of Kraemmer and Rebhan is 
{\it wrong\/}.  We therefore propose a new proof of the KR proposition, which
requires the detailed analysis of the conformal-anomaly term.
Our analysis clarifies under what circumstances the KR proposition is valid.
There is no contradiction between the KR proposition and our claim.
This point will be discussed in detail in a separate paper \cite{AN2}.
\par
The present paper is organized as follows. In Section 2, we briefly review
the KR proposition and its proof of Kraemmer and Rebhan critically.
In Section 3, we systematize the calculation of the conformal-anomaly term.
In this connection, it is found convenient that the terms appearing in 
$T_{\mu\nu}$ are classified into several ``types''. 
In Section 4, we consider deformations of the gauge-fixing and FP-ghost
Lagrangian density under certain conditions.  We then find that under
deformations there are four conservation laws for the numbers of terms 
characterized by the above-mentioned ``types''.  From those conservation
laws, the invariance of the conformal-anomaly term under deformations follows.
The final section is devoted to discussions, where some remarks are made on
the anomaly of the FP-ghost number current.
\vskip50pt
%
%
\Sec{Kraemmer-Rebhan's work}
In this section, we review the work of Kraemmer and 
Rebhan \cite{Kraemmer-Rebhan} concerning the gauge independence of the 
conformal anomaly.
\par
The covariant-gauge two-dimensional quantum gravity is described by the 
gravitational field $g_{\mu\nu}$, the B-field $b_\rho$, the FP ghost 
$c^\sigma$ and antighost $\bar c_\tau$, together with scalar fields $X_M$ $(M=
0,\,1,\,\ldots,\,D-1)$.  We write $g\equiv\det g_{\mu\nu}$ and $\tg^{\mu\nu}
\equiv\sqrt{-g}g^{\mu\nu}$.  The BRS transformation of those fields are as 
follows:
\begin{eqnarray}
&&\gdel g_{\mu\nu}=-\partial_\mu c^\sigma\cdot g_{\sigma\nu}
                 -\partial_\nu c^\sigma\cdot g_{\mu\sigma}
                 -c^\sigma\partial_\sigma g_{\mu\nu}, \\
&&\gdel X_M=-c^\lambda\partial_\lambda X_M, \\
&&\gdel c^\sigma=-c^\lambda\partial_\lambda c^\sigma, \\
&&\gdel \bar c_\tau=ib_\tau, \\
&&\gdel b_\rho=0.
\end{eqnarray}
In general, we should also consider the Weyl BRS transformation, but since
the conformal degree of freedom is eliminated, we omit describing 
the Weyl gauge-fixing plus Weyl FP-ghost Lagrangian density for simplicity.
\par
The total Lagrangian density $\lag$ consists of the string one,
\begin{eqnarray}
&&\lag_{\hbox{\sc S}}=
{1\over2}\tg^{\mu\nu}\partial_\mu X^M\cdot\partial_\nu X_M,
\end{eqnarray}
the gauge-fixing one, $\lag_{\hbox{\sc GF}}$, and the FP-ghost one, 
$\lag_{\hbox{\sc FP}}$.   Of course, $\lag_{\hbox{\sc S}}$ and 
$\lag_{\hbox{\sc GF}} + \lag_{\hbox{\sc FP}}$ are BRS-invariant up to 
total-divergence terms.
\par
Kraemmer and Rebhan \cite{Kraemmer-Rebhan} employ the perturbative approach
{\it in the intrinsic way\/}, that is, they employ it {\it before\/} writing
down the expression for $\lag_{\hbox{\sc GF}} + \lag_{\hbox{\sc FP}}$.
They first set
\begin{eqnarray}
&&g_{\mu\nu}=\hat g_{\mu\nu}+h_{\mu\nu},
\end{eqnarray}
where $\hat g_{\mu\nu}$ denotes the classical background metric, while
$h_{\mu\nu}$ does the ``small'' quantum gravitational field.
After substituting \eqno(2,7) into $\lag$, only terms quadratic in quantum 
fields are retained.  Since this model has no higher-loop corrections, it is
perturbatively justifiable to neglect higher-order terms.
It is worth noting that Kraemmer and Rebhan also neglected linear terms.
The reason for this is that linear terms do not contribute to the conformal
anomaly in the perturbative calculation.
\par
Since $\lag_{\hbox{\sc GF}}$ is proportional to $b_\rho$, the above 
prescription implies that $\lag_{\hbox{\sc GF}}$ is also proportional to 
$h_{\mu\nu}$, that is, we have
\begin{eqnarray}
&&\lag_{\hbox{\sc GF}}=b_\rho F^{\rho\sigma\tau}h_{\sigma\tau},
\end{eqnarray}
where $F^{\rho\sigma\tau}$ is a first-order differential operator depending on
$\hat g_{\mu\nu}$ and its derivatives.   The condition that the conformal
degree of freedom is eliminated is expressed as
\begin{eqnarray}
&&F^{\rho\sigma\tau}\hat g_{\sigma\tau}=0.
\end{eqnarray}
\par
The FP-ghost Lagrangian density is determined by
\begin{eqnarray}
&&\lag_{\hbox{\sc GF}} + \lag_{\hbox{\sc FP}}=
 -i\gdel^{(0)}(\bar c_\rho F^{\rho\sigma\tau}h_{\sigma\tau}).
\end{eqnarray}
Here $\gdel^{(0)}$ is the linearized BRS transformation obtained 
from \eqno(2,1)--\eqno(2,5) by substituting \eqno(2,7) and 
by neglecting nonlinear terms in quantum fields.  We find
\begin{eqnarray}
\gdel^{(0)}h_{\mu\nu}
&=&-\hat g_{\mu\sigma}\partial_\nu c^\sigma 
   -\hat g_{\sigma\nu}\partial_\mu c^\sigma
   -\partial_\sigma\hat g_{\mu\nu}\cdot c^\sigma \nonumber \\
&=&-(\hat\nabla_\mu c_\nu + \hat\nabla_\nu c_\mu),\\
\gdel^{(0)}X_M&=&0,\\
\gdel^{(0)}c^\sigma&=&0,\\
\gdel^{(0)}\bar c_\tau&=&ib_\tau,\\
\gdel^{(0)}b_\rho&=&0,
\end{eqnarray}
where $\hat\nabla_\lambda$ denotes covariant differentiation with respect to
$\hat g_{\mu\nu}$.  Here the raising and lowering of the index of $c$ are made
by using $\hat g^{\mu\nu}$ and $\hat g_{\mu\nu}$.  Because of the appearance
of covariant differentiation, the expression \eqno(2,11) is essentially 
irrelevant to whether $c^\sigma$ or $c_\sigma$ is chosen as the primary field.
Substituting \eqno(2,11) into \eqno(2,10), we obtain
\begin{eqnarray}
\lag_{\hbox{\sc FP}}=-i\bar c_\rho F^{\rho\sigma\tau}
                     (\hat\nabla_\sigma c_\tau
                      +\hat\nabla_\tau c_\sigma).
\end{eqnarray}
\par
In order to avoid the appearance of second-order derivatives in \eqno(2,16), it
is usual to make partial integration in \eqno(2.8).  We then have
\begin{eqnarray}
&&\lag_{\hbox{\sc GF}}=(F_1{}^{\lambda\rho\sigma\tau}\partial_\lambda b_\rho
 + F_2{}^{\rho\sigma\tau}b_\rho)h_{\sigma\tau}, \\
&&\lag_{\hbox{\sc FP}}=-i(
 F_1{}^{\lambda\rho\sigma\tau}\partial_\lambda\bar c_\rho
+F_2{}^{\rho\sigma\tau}\bar c_\rho)(\hat\nabla_\sigma c_\tau
                                   +\hat\nabla_\tau c_\sigma),
\end{eqnarray}
which are used in perturbative  calculation.
\par
Now, the conformal anomaly is discussed on the basis of calculating
\begin{eqnarray}
&&{\delta^2\Gamma\over\delta\hat g_{\mu\nu}\delta\hat g_{\lambda\rho}},
\end{eqnarray}
where $\Gamma$ is the effective action.  This is equivalent to calculating
the two-point function of $T_{\mu\nu}$, i.e.,
\begin{eqnarray}
&&\wightman{T_{\mu\nu}(x)\,T_{\lambda\rho}(y)}, 
\end{eqnarray}
where
\begin{eqnarray}
&&T_{\mu\nu}\equiv2(-\hat g)^{-1/2}{\delta\over\delta\hat g^{\mu\nu}}
\int d^2 x\,\lag
\end{eqnarray}
and $\wightman{\ \cdots\ }$ denotes the vacuum expectation value of 
a time-ordered product.  Usually, we take the flat limit 
$\hat g_{\mu\nu}=\eta_{\mu\nu}$ in \eqno(2,20) to avoid unnecessary 
complication.   $T_{\mu\nu}$ is the symmetric energy-momentum tensor if
$\lag$ is a scalar density under general coordinate transformation with 
respect to the background metric [1, 5--7]
Kraemmer and Rebhan \cite{Kraemmer-Rebhan}, however, do not impose this
condition, whence $T_{\mu\nu}$ is no longer a conserved quantity in 
general.
\par
The two-point function \eqno(2,20) is calculated by means of Feynman rules.
If \eqno(2,8) with $\hat g_{\mu\nu}=\eta_{\mu\nu}$ reproduces the homogeneous
linear part of the de Donder gauge fixing, the nonvanishing propagators are
given by 
\begin{eqnarray}
&&\wightman{X_M(x)X_N(y)}=\eta_{MN}\Df(x-y),\\
&&\wightman{h_{\mu\nu}(x)b_\rho(y)}=
 (\eta_{\mu\rho}\partial_\nu+\eta_{\rho\nu}\partial_\mu
  -\eta_{\mu\nu}\partial_\rho)^x\Df(x-y),\\
&&\wightman{c^\sigma(x)\bar c_\tau(y)}=i\delta^\sigma{}_\tau\Df(x-y).
\end{eqnarray}
One-loop integrals are evaluated with the help of dimensional regularization.
The result can be written as
\begin{eqnarray}
&&\wightman{T_{\mu\nu}(x)T_{\lambda\rho}(y)}
=(D+a)\vPhi_{\mu\nu\lambda\rho}(x-y)+\hbox{local terms},
\end{eqnarray}
where $a$ is a numerical constant and the Fourier transform of 
$\vPhi_{\mu\nu\lambda\rho}$ is proportional to $p_\mu p_\nu p_\lambda p_\rho /
(p^2+i0)$.  The first term of the r.h.s.\ of \eqno(2,25) is the 
conformal-anomaly term.  It vanishes if and only if $D+a=0$.
Many concrete examples yield $a=-26$.
\par
The KR proposition is as follows.
\vskip3pt
\par
{\it The value of $a$ is independent of the choice of $F^{\rho\sigma\tau}$.}
\vskip3pt
\par
The proof of Kraemmer and Rebhan \cite{Kraemmer-Rebhan} is as follows.
Let $\Gamma^{(1)}$ be the one-loop contribution to $\Gamma$.  Consider any
infinitesimal variation $\delta F^{\rho\sigma\tau}$ of $F^{\rho\sigma\tau}$.
Then we have
\begin{eqnarray}
\delta\Gamma^{(1)}[\hat g]&=&-i\wightman{\gdel^{(0)}
\Big(\int d^2x\, \bar c_\rho
 \delta F^{\rho\sigma\tau}h_{\sigma\tau}\Big)} \nonumber\\
&=&0
\end{eqnarray}
because the BRS invariance is not violated spontaneously.
From \eqno(2,26), they inferred that $\delta a=0$.
\par
If their proof were correct, then the same reasoning for $F^{\rho\sigma\tau}$
would imply that $a=0$.  Evidently, something must be wrong.  
Indeed, in order to deduce the invariance of \eqno(2,20) from \eqno(2,26), 
{\it we need the commutativity between $\gdel^{(0)}$ and 
$\delta/\delta\hat g^{\mu\nu}$, but they actually do not commute.\/}
In fact, from \eqno(2,11) we have
\begin{eqnarray}
&&{\delta\over\delta\hat g^{\mu\nu}}\Big(\gdel^{(0)}h_{\lambda\rho}\Big)\not=0=
\gdel^{(0)}\Big({\delta\over\delta\hat g^{\mu\nu}}h_{\lambda\rho}\Big).
\end{eqnarray}
Thus we can say nothing about $\wightman{T_{\mu\nu}T_{\lambda\rho}}$
from \eqno(2,26).
\par
To prove the KR proposition, we must make the detailed analysis of 
$\wightman{T_{\mu\nu}T_{\lambda\rho}}$ itself.
%
%
\Sec{Classification of the terms of $T_{\mu\nu}$}
Before entering into the general consideration, we consider a concrete example
for better understanding.
\par
The simplest choice of gauge fixing which satisfies \eqno(2,9) and reduces
to the linearized de Donder gauge for $\hat g_{\mu\nu}=\eta_{\mu\nu}$ is 
\begin{eqnarray}
&&\lagGF=\partial^\sigma b^\tau\cdot(h_{\sigma\tau}
 -{1\over2}\hat g_{\sigma\tau}\hat g^{\alpha\beta}h_{\alpha\beta}),\\
\noalign{\noindent where $b^\tau=\eta^{\tau\rho}b_\rho$, and 
from \eqno(2,18) the corresponding $\lagFP$ is}
&&\lagFP=-i\partial^\sigma\bar c^\tau\cdot(\hat\nabla_\sigma c_\tau
 +\hat\nabla_\tau c_\sigma
 -\hat g_{\sigma\tau}\hat g^{\alpha\beta}\hat\nabla_\alpha c_\beta)
\end{eqnarray}
with $\bar c^\tau=\eta^{\tau\rho}\bar c_\rho$.
Then \eqno(2,21) with $\hat g_{\mu\nu}=\eta_{\mu\nu}$ yields
\def\TS{T^{\hbox{\sc S}}{}}
\def\TGF{T^{\hbox{\sc GF}}{}}
\def\TFP{T^{\hbox{\sc FP}}{}}
\begin{eqnarray}
&&T_{\mu\nu}=\TS_{\mu\nu}+\TGF_{\mu\nu}+\TFP_{\mu\nu}, \\
&&\TS_{\mu\nu}=\partial_\mu X^M\cdot\partial_\nu X_M+\cdots, \\
&&\TGF_{\mu\nu}=-\partial^\sigma b_\sigma\cdot h_{\mu\nu} + \cdots, \\
&&i\TFP_{\mu\nu}=[2\partial_\mu\bar c_\nu\cdot\partial_\sigma c^\sigma
 +\partial_\sigma\partial_\mu\bar c_\nu\cdot c^\sigma
 -\partial_\mu\bar c_\sigma\cdot\partial^\sigma c_\nu 
 -\partial_\sigma\bar c_\mu\cdot\partial^\sigma c_\nu
\nonumber\\ &&\hspace{70pt} 
  + \ (\mu\leftrightarrow\nu)\ ] \ +\ \cdots,
\end{eqnarray}
where dots indicate terms proportional to $\eta_{\mu\nu}$, which are 
irrelevant to the conformal anomaly.  In \eqno(3,5), the free-field
equation for $b_\rho$,
\begin{eqnarray}
&&\partial_\lambda b_\rho +\partial_\rho b_\lambda 
 - \eta_{\lambda\rho}\partial^\sigma b_\sigma=0,
\end{eqnarray}
has been used.
\par
The two-point function,
\begin{eqnarray}
&&\wightman{T_{\mu\nu}T_{\lambda\rho}}=
\wightman{\TS_{\mu\nu}\TS_{\lambda\rho}}
+\wightman{\TGF_{\mu\nu}\TGF_{\lambda\rho}}
+\wightman{\TFP_{\mu\nu}\TFP_{\lambda\rho}},
\end{eqnarray}
is calculated by using \eqno(2,22)--\eqno(2,24) and the formulae
\begin{eqnarray}
&&\partial_\mu\partial_\nu\Df\cdot\partial_\lambda\partial_\rho\Df
 ={1\over2}\vPhi_{\mu\nu\lambda\rho}+\hbox{local terms}, \\
&&\partial_\mu\partial_\nu\partial_\lambda\Df\cdot\partial_\rho\Df
 =\vPhi_{\mu\nu\lambda\rho}+\hbox{local terms}.
\end{eqnarray}
It is easy to see that the contribution to $a$ from 
$\wightman{\TGF_{\mu\nu}\TGF_{\lambda\rho}}$ is $+2$, while that from
$\wightman{\TFP_{\mu\nu}\TFP_{\lambda\rho}}$ is $-28$; hence $a=2-28=-26$.
\par
It is interesting to see how $-28$ results from 
$\wightman{\TFP_{\mu\nu}\TFP_{\lambda\rho}}$.
We symbolically write $\partial_\mu\bar c_\nu\cdot\partial_\sigma c^\sigma
+(\mu\leftrightarrow\nu)$ as $I$, $\partial_\sigma\partial_\mu\bar c_\nu\cdot
c^\sigma + (\mu\leftrightarrow\nu)$ as $I\!I$, and so on, so that the r.h.s.\ 
of \eqno(3,6) is symbolically written as $2I+I\!I-I\!I\!I-I\!V$.
The contribution to $a$ is given by 
$2^2\wightman{I\,I}+2\cdot2\wightman{I\,I\!I}
+\wightman{I\!I\,I\!I}+\wightman{I\!I\!I\,I\!I\!I}\sim2^2(-2)
+2\cdot2(-4)+(-2)+(-2)=-28$.
It is important to note that $\wightman{I\,I\!I\!I}\sim0$, 
$\wightman{I\!I\,I\!I\!I}\sim0$ and $IV$ does not contribute at all. 
Those facts come from whether or not each of $c$ and $\bar c$ has an external
index $\mu$ or $\nu$.
\par
Now, we consider the general situation.  From the above remark, it is 
reasonable to classify the terms appearing in $i\TFP_{\mu\nu}$ into the
following four ``types''.
\begin{eqnarray*}
\hbox{Type $AB$}&:&\hbox{Neither $c$ nor $\bar c$ have an external index.} \\
\hbox{Type $A$}&:& \hbox{Only $c$ has an external index.}\\
\hbox{Type $B$}&:& \hbox{Only $\bar c$ has an external index.}\\
\hbox{Type $O$}&:& \hbox{Both $c$ and $\bar c$ have an external index.}
\end{eqnarray*}
Furthermore, we call the term having first-order derivatives only $Rh+$ and
the one having a second-order derivative $Rh-$.
We abbreviate (Type $A$, $Rh+$) as $A^+$, and so on.  For example, for the
terms in \eqno(3,6), we see that $I \sim B^+$, $I\!I \sim B^-$, $I\!I\!I 
\sim A^+$, $I\!V \sim O^+$.   Note that $2B^+ + B^-$ yields $-26$ precisely.
\par
We list up the possible terms appearing in $T_{\mu\nu}$ and classify them 
according to the above classification.
\vskip10pt
\par
\begin{center}
\begin{tabular}{c|l|ll}
      & \hfil $Rh+$ \hfil & \multicolumn{2}{c}{$Rh-$} \\
\hline
$AB$  & $\partial_\mu\bar c_\sigma\cdot\partial_\nu c^\sigma 
        + (\mu\leftrightarrow\nu)$
      & $2\partial_\mu\partial_\nu\bar c_\sigma\cdot c^\sigma$,
      & $2\bar c_\sigma\partial_\mu\partial_\nu c^\sigma$ \\
\hline
$A$   & $\partial_\mu\bar c_\sigma\cdot\partial^\sigma c_\nu
        + (\mu\leftrightarrow\nu)$ 
      & \multicolumn{2}{l}{$\partial_\mu\partial^\sigma\bar c_\sigma\cdot c_\nu
        + (\mu\leftrightarrow\nu)$}  \\
      & $\partial^\sigma\bar c_\sigma\cdot\partial_\mu c_\nu
        + (\mu\leftrightarrow\nu)$
      & 
      & $\bar c_\sigma\partial^\sigma\partial_\mu c_\nu
        + (\mu\leftrightarrow\nu)$ \\
\hline
$B$   & $\partial_\sigma\bar c_\mu\cdot\partial_\nu c^\sigma
        + (\mu\leftrightarrow\nu)$ 
      & \multicolumn{2}{l}{$\partial_\nu\partial_\sigma c_\mu\cdot c^\sigma
        + (\mu\leftrightarrow\nu)$} \\
      & $\partial_\mu\bar c_\nu\cdot\partial^\sigma c_\sigma
        + (\mu\leftrightarrow\nu)$
      &
      & $\bar c_\nu\partial_\mu\partial_\sigma c^\sigma
        + (\mu\leftrightarrow\nu)$ \\
\hline
$O$   & $\partial_\sigma\bar c_\mu\cdot\partial^\sigma c_\nu
        + (\mu\leftrightarrow\nu)$
      & \multicolumn{2}{c}{---}
\end{tabular}
\end{center}
\par
Some remarks are in order.  $O^-$ is nonexistent because $\square c_\sigma=
\square\bar c_\sigma=0$.  Type $O$ is effectively equivalent to zero.
The cross term of Type $A$ and Type $B$ vanishes.  Type $AB$ is effectively
equivalent to Type $A$ plus Type $B$.
\par
A special care is necessary for $Rh$ $-$.  Since $\partial_\mu\partial_\nu
\partial_\lambda\partial_\rho\Df\cdot\Df$ is infrared divergent, we must 
check the nonappearance of nonvanishing cross terms of the form 
$\wightman{(\partial\partial\bar c\cdot c)(\bar c\partial\partial c)}$.
By explicit calculation based on the FP-ghost Lagrangian density \eqno(4,5) 
in the next section, it is found that we encounter only the following
three kinds of the $Rh-$ terms.
\begin{eqnarray}
&&A^- \; : \qquad \bar c_\sigma\partial^\sigma\partial_\mu c_\nu
                  + (\mu\leftrightarrow\nu), \nonumber \\
&&B^- \; : \qquad \partial_\nu\partial_\sigma\bar c_\mu\cdot c^\sigma
                  + (\mu\leftrightarrow\nu), \nonumber \\
&&A^- \; : \qquad (\bar c_\sigma\partial_\mu\partial_\nu c^\sigma
                  -\bar c_\nu\partial_\mu\partial_\sigma c^\sigma)
                  + (\mu\leftrightarrow\nu). \nonumber
\end{eqnarray}
The last one is Type $A$ because it is Type $AB$ minus Type $B$.
Since the cross term of Type $A$ and Type $B$ vanish, we do not encounter
infrared divergence.
\par
Next, we classify the terms appearing in $\TGF_{\mu\nu}$ in the following way.
\begin{eqnarray*}
\hbox{Type $A$}&:& \hbox{$b$ has no external index, }\\
\hbox{Type $O$}&:& \hbox{$b$ has an external index;}\\
Rh+ &:& \hbox{Differential operator acts on $b$,}\\
Rh- &:& \hbox{Differential operator acts on $h$.}
\end{eqnarray*}
We encounter the following terms.
\vskip10pt
\begin{center}
\begin{tabular}{c|c|c}
         & $Rh+$ & $Rh-$ \\
\hline
Type $A$ & ${1\over2}[\partial_\mu b_\sigma\cdot h_\nu{}^\sigma
           + (\mu\leftrightarrow\nu)]$
         & ${1\over2}[b_\sigma\partial_\mu h_\nu{}^\sigma 
           + (\mu\leftrightarrow\nu)]$ \\
         & $\partial^\sigma b_\sigma\cdot h_{\mu\nu}$
         & $b_\sigma\partial^\sigma h_{\mu\nu}$ \\
\hline
Type $O$ & ${1\over2}[\partial_\sigma b_\mu\cdot h_\nu{}^\sigma
           + (\mu\leftrightarrow\nu)]$
         & ${1\over2}[b_\mu\partial_\sigma h_\nu{}^\sigma
           + (\mu\leftrightarrow\nu)]$
\end{tabular}
\end{center}
\vskip10pt\noindent
Type $O$ is effectively equivalent to zero.  The two terms of $A^+$ are
effectively equal because their difference is of Type $O$ owing to \eqno(3,7).
\par
We denote the algebraic sum of the coefficients of the $A^\pm$ and 
$AB^\pm$ terms appearing in $i\TFP_{\mu\nu}$ by $N(A^\pm)$, that of the
$B^\pm$ and $AB^\pm$ terms appearing in $i\TFP_{\mu\nu}$ by $N(B^\pm)$
and that of the $A^\pm$ terms appearing in $\TGF_{\mu\nu}$ by $M(A^\pm)$.
Then we have the general formula
\begin{eqnarray}
a&=&[M(A^+)]^2\times2+2M(A^+)M(A^-)\times4+[M(A^-)]^2\times2 \nonumber\\
 && +[N(A^+)]^2\times(-2)+2N(A^+)N(A^-)\times(-4)+[N(A^-)]^2\times(-2)
 \nonumber \\
 && +[N(B^+)]^2\times(-2)+2N(B^+)N(B^-)\times(-4)+[N(B^-)]^2\times(-2).
\end{eqnarray}
Therefore, if
\begin{eqnarray}
&&N(A^+)=M(A^+), \\
&&N(A^-)=M(A^-), \\
&&N(B^+)=2, \\
&&N(B^-)=1,
\end{eqnarray}
then we obtain $a=-26$.
\vskip30pt
%
%
\Sec{Deformations and conservation laws}
We discuss the general case of $\lagGF+\lagFP$ given by \eqno(2,17) and
\eqno(2,18).  We require formal tensorial invariance.  Then we may assume,
without loss of generality, that $F_1{}^{\lambda\rho\sigma\tau}$ involves
no $\partial_\kappa\hat g_{\alpha\beta}$ and that $F_2{}^{\rho\sigma\tau}$ is
proportional to $\partial_\kappa\hat g_{\alpha\beta}$ because we need 
$T_{\mu\nu}$ only for $\hat g_{\mu\nu}=\eta_{\mu\nu}$.  
We rewrite \eqno(2,17) into the form which manifestly exhibits the satisfaction
of the condition \eqno(2,9) and reduces to the linearized de Donder gauge
fixing in the flat limit:
\begin{eqnarray}
&&\lagGF=\sum_{j=1}^2(\tilde F_{1j}{}^{\lambda\rho\sigma\tau}[\hat g]
 \partial_\lambda b_\rho +\tilde F_{2j}{}^{\rho\sigma\tau}[\hat g,\,\partial
\hat g]b_\rho)H^j{}_{\sigma\tau},\\
\noalign{\noindent where}
&&H^1{}_{\sigma\tau}\equiv h_{\sigma\tau}
 -{1\over2}\hat g_{\sigma\tau}\hat g^{\alpha\beta}h_{\alpha\beta}, \\
&&H^2{}_{\sigma\tau}\equiv h_{\sigma\tau}
 -\hat g_{\sigma\tau}{\eta^{\alpha\beta}h_{\alpha\beta}\over
                      \eta^{\gamma\delta}\hat g_{\gamma\delta}},\\
\noalign{\noindent and}
&&\sum_{j=1}^2\tilde F_{1j}{}^{\lambda\rho\sigma\tau}[\eta]
 =\eta^{\lambda\sigma}\eta^{\rho\tau}.
\end{eqnarray}
Of course, $\tilde F_{2j}{}^{\rho\sigma\tau}=0$ for 
$\hat g_{\mu\nu}=\eta_{\mu\nu}$.
Correspondingly, \eqno(2,18) is rewritten as
\begin{eqnarray}
&&\lagFP=-i\sum_{j=1}^2(\tilde F_{1j}{}^{\lambda\rho\sigma\tau}[\hat g]
\partial_\lambda\bar c_\rho
 +\tilde F_{2j}{}^{\rho\sigma\tau}[\hat g,\partial\hat g]\bar c_\rho)
 K^j{}_{\sigma\tau},\\
\noalign{\noindent where}
&&K^1{}_{\sigma\tau}\equiv\hat\nabla_\sigma c_\tau+\hat\nabla_\tau c_\sigma
      -\hat g_{\sigma\tau}\hat g^{\alpha\beta}\hat\nabla_\alpha c_\beta, \\
&&K^2{}_{\sigma\tau}\equiv\hat\nabla_\sigma c_\tau+\hat\nabla_\tau c_\sigma
      -\hat g_{\sigma\tau}{2\eta^{\alpha\beta}\hat\nabla_\alpha c_\beta\over
       \eta^{\gamma\delta}\hat g_{\gamma\delta}}.
\end{eqnarray}
\par
We start with the two simplest cases:
\begin{eqnarray}
\hspace*{-80pt}
\hbox{(i)}&&\tilde F_{11}{}^{\lambda\rho\sigma\tau}
=\eta^{\lambda\sigma}\eta^{\rho\tau}, \qquad \tilde F_{12}=\tilde F_{2j}=0; \\
\hspace*{-80pt}
\hbox{(ii)}&&\tilde F_{12}{}^{\lambda\rho\sigma\tau}
=\eta^{\lambda\sigma}\eta^{\rho\tau}, \qquad \tilde F_{11}=\tilde F_{2j}=0.
\end{eqnarray}
The first case \eqno(4,8) is nothing but the example discussed at the beginning
of Section 3.   From \eqno(3,5) and \eqno(3,6), we find that
\begin{eqnarray}
&&M(A^+)=-1, \qquad M(A^-)=0; \nonumber \\
&&N(A^+)=-1, \qquad N(A^-)=0, \qquad N(B^+)=2, \qquad N(B^-)=1.
\end{eqnarray}
In the second case \eqno(4,9), we have
\begin{eqnarray}
&&\TGF_{\mu\nu}=0+\cdots,\\
&&i\TFP_{\mu\nu}=[2\partial_\mu\bar c_\nu\cdot\partial_\sigma c^\sigma
+\partial_\sigma\partial_\mu\bar c_\nu\cdot c^\sigma 
+\partial^\sigma\bar c_\sigma\cdot\partial_\mu c_\nu
-\partial_\mu\bar c_\sigma\cdot\partial^\sigma c_\nu
-\partial_\sigma\bar c_\mu\cdot\partial^\sigma c_\nu \nonumber \\
&&\hspace*{70pt} +\ (\mu\leftrightarrow\nu)\ ]\ +\ \cdots,
\end{eqnarray}
where use has been made of \eqno(3,7).  Accordingly, 
\begin{eqnarray}
&&M(A^+)=M(A^-)=0; \nonumber \\
&&N(A^+)=N(A^-)=0, \qquad N(B^+)=2, \qquad N(B^-)=1.
\end{eqnarray}
In both cases, \eqno(3,12)--\eqno(3,15) hold. 
\par
Now, we proceed to the general case.  To do this, we employ {\it finite 
deformations\/} rather than infinitesimal variations.  
By combining the following three kinds of deformations, we can achieve the 
general expression \eqno(4,1).
\vskip10pt
\begin{item1}
\item[\bf D1.] To take a linear combination of any two already constructed 
gauge fixings in such a way that the normalization condition \eqno(4,4) is
kept.
\item[\bf D2.] To insert a background metric, that is, to replace 
$\delta^\alpha{}_\beta$ by $\hat g^{\alpha\gamma}\eta_{\gamma\beta}$ or
$\eta^{\alpha\gamma}\hat g_{\gamma\beta}$.\foot(3,{Replacement of 
$\delta^\alpha{}_\beta$ by $\eta^{\alpha\gamma}\hat g_{\gamma\beta}$ in 
{$f(\det \delta^\alpha{}_\beta)$} implies the insertion of an arbitrary
function of $-\hat g$.  Of course, such insertion is irrelevant to the 
conformal anomaly.})
\item[\bf D3.] To add a term proportional to 
$\partial_\kappa\hat g_{\alpha\beta}\cdot b_\rho H^j{}_{\sigma\tau}$.
\end{item1}
\vskip10pt
\par
As for {\bf D1}, it is self-evident that \eqno(3,12)--\eqno(3,15) hold if 
they hold for both constituents.  As for {\bf D2} and {\bf D3}, we have only to
take account of the new contribution to $T_{\mu\nu}$ due to the deformation
considered, and therefore the other parts of $\lagGF+\lagFP$ may be reduced to
their expressions in the flat limit {\it beforehand\/}.
\par
First, we consider the Type-$A$ conservation laws \eqno(3,12) and \eqno(3,13).
They are nothing but the consequence of the BRS invariance.  We observe:
\begin{item1}
\item[1.] The deformation made in $\lagGF$ is exactly the same as in the one
made in $\lagFP$ if $b_\rho$ is replaced by $\bar c_\rho$.
\item[2.] If $b$ has an internal index, $\bar c$ has also an internal index,
and {\it vice versa\/}.  Thus Type $A$ in $\TGF_{\mu\nu}$ corresponds to Type
$A$ or Type $AB$ in $i\TFP_{\mu\nu}$, while Type $O$ in $\TGF_{\mu\nu}$ 
corresponds to Type $O$ or Type $B$ in $i\TFP_{\mu\nu}$.
\item[3.] The distinction between $Rh+$ and $Rh-$ is common in $\TGF_{\mu\nu}$
and in $i\TFP_{\mu\nu}$.
\end{item1}
From these properties, we can conclude that the change of $M(A^\pm)$ due to 
the deformation is equal to that of $N(A^\pm)$.
\par
Next, we proceed to the Type-$B$ conservation laws \eqno(3,14) and \eqno(3,15).
In this case, of course, we have only to consider $i\TFP_{\mu\nu}$.
In the flat background, both $K^1{}_{\sigma\tau}$ and $K^2{}_{\sigma\tau}$
reduce to
\begin{eqnarray}
&&\partial_\tau c_\sigma+\partial_\sigma c_\tau
   -\eta_{\sigma\tau}\partial^\alpha c_\alpha.
\end{eqnarray}
Type $B$ and Type $AB$ are characterized by the property that $c$ has an 
internal index.   There are four possibilities in whether or not $\sigma$ and
$\tau$ are replaced by $\mu$ and $\nu$ in the new terms arisen in 
$i\TFP_{\mu\nu}$ owing to the deformation.
\begin{item1}
\item[1.] The case in which neither $\sigma$ nor $\tau$ is replaced by an 
external index.  Then since both $\sigma$ and $\tau$ are internal, they must
be contracted in $i\TFP_{\mu\nu}$.  Hence we have
\begin{eqnarray}
&&\eta^{\sigma\tau}(\partial_\tau c_\sigma+\partial_\sigma c_\tau
 -\eta_{\sigma\tau}\partial^\alpha c_\alpha)=0,
\end{eqnarray}
that is, there is no change in $N(B^\pm)$.
\item[2.] The case in which $\sigma$ alone is replaced by an external index
$\mu$ or $\nu$.  Then the first term of \eqno(4,14) gives no change to 
$N(B^\pm)$.  The changes due to the second term and the third one just cancel
out.
\item[3.] The case in which $\tau$ alone is replaced by $\mu$ or $\nu$.
Similar to Case 2.
\item[4.] The case in which both $\sigma$ and $\tau$ are replaced by external
indices.  Then, evidently, neither the first term of \eqno(4,14) nor the 
second term contributes to the change of $N(B^\pm)$.  Furthermore, since the
third term is proportional to $\eta_{\mu\nu}$, it does not contribute.
\end{item1}
\par
Thus, in all cases, $N(B^\pm)$ remains unchanged.  Accordingly, \eqno(3,14) and
\eqno(3,15) hold always.
\par
From the above consideration, we have seen that \eqno(3,12)--\eqno(3,15) are
valid for any $\TGF_{\mu\nu}$ and $i\TFP_{\mu\nu}$ obtainable from 
\eqno(4,1) and \eqno(4,5).  This completes the proof of the KR proposition.
\vskip30pt
%
%
%
%
\Sec{Discussions}
\def\jc{j_{\hbox{\sc c}}{}}
\def\jL{j_{\hbox{\sc L}}{}}
\def\ac{a_{\hbox{\sc c}}}
\def\aL{a_{\hbox{\sc L}}}
In the present paper, we have succeeded in proving the gauge independence of
the conformal anomaly in the sense of Kraemmer and Rebhan.  It is important 
to note that the BRS invariance is relevant to the terms of Type $A$ only.
{\it The critical value $a=-26$ arises from the terms of Type $B$\/}.
Its invariance is due to the special properties of the particular expression 
\eqno(4,14) rather than the BRS invariance.
\par
Although the original form \cite{Kraemmer-Rebhan} of the KR proposition 
includes also noncovariant gauges, our proof applied only  to the gauges 
which  reduce to the linearized de Donder gauge in the flat limit.  
This is because we should not change the Feynman rules.   
Hence we cannot yet give a satisfactory answer to the question on 
whether it is inevitable or accidental that both
covariant gauges and noncovariant gauges give $a$ the same value $-26$.
As will be discussed in detail in a separate paper \cite{AN2}, the validity of
the KR proposition is dependent on the perturbative approach {\it in the 
crucial way\/}.  Hence it is quite difficult to compare two gauges which have
different Feynman rules.
\par
Kraemmer and Rebhan \cite{Kraemmer-Rebhan} presented also a proof of the gauge
independence of the anomaly of the so-called ``Rebhan-Kraemmer current''
\cite{Kummer, PSS}, which is a sum of the FP-ghost number current $\jc^\mu$
and the Lagrangian-multiplier (or $b$-field) current $\jL^\mu$.
However, as long as one considers the two-point functions $\wightman{\jc^\mu
T_{\lambda\rho}}$ and $\wightman{\jL^\mu T_{\lambda\rho}}$, our criticism
applies also to the proof of the gauge independence of the anomaly of the 
Rebhan-Kraemmer current.
\par
Rebhan and Kraemmer \cite{Rebhan-Kraemmer} emphasized the difference between 
the expressions for $\jc^\mu$ in D\"usedau's case (background generally
covariant) \cite{Dusedau} and in their case (given by \eqno(4,9)).
However, the {\it Noether currents\/} in both cases coincide in the flat limit:
\begin{eqnarray}
&&i \jc^\mu=\bar c_\sigma\partial^\mu c^\sigma-\partial^\mu\bar c_\sigma\cdot
c^\sigma+\partial_\sigma(\bar c^\sigma c^\mu-\bar c^\mu c^\sigma),
\end{eqnarray}
as it should be.  D\"usedau \cite{Dusedau} neglected the total-divergence terms
tacitly, but {\it this negligence is the origin of the difference\/}.
As emphasized in Ref.\ \cite{AN1}, {\it we must not neglect total-divergence
terms in the anomaly calculation\/}.  The total-divergence terms of \eqno(5,1)
actually contribute to the anomaly. Indeed, if Rebhan and Kraemmer also had 
unconsciously neglected the total-divergence terms, they would have been led
to a different conclusion.
\par
The anomaly of $\wightman{\jc^\mu\,T_{\lambda\rho}}$ arises from the loop
integral $\partial\partial\Df\cdot\partial\Df$.  The general formula for the
coefficient is 
\begin{eqnarray}
&&\ac=N(A^+)+N(A^-)-N(B^+)-N(B^-).
\end{eqnarray}
On the other hand, since
\begin{eqnarray}
&&\jL^\mu=-b_\sigma h^{\mu\sigma}+{1\over2}b^\mu h^\sigma{}_\sigma
\end{eqnarray}
in the flat limit, we have
\begin{eqnarray}
&&\aL=-M(A^+)-M(A^-).
\end{eqnarray}
From \eqno(3,12)--\eqno(3,15), therefore, we find
\begin{eqnarray}
&&\ac+\aL=-3.
\end{eqnarray}
This establishes the gauge independence of the anomaly of the Rebhan-Kraemmer
current.
\vfill\eject
%
%

\end{document}